\documentclass[fleqn,usenatbib]{mnras}

\usepackage{newtxtext,newtxmath}

\usepackage[T1]{fontenc}

\DeclareRobustCommand{\VAN}[3]{#2}
\let\VANthebibliography\thebibliography
\def\thebibliography{\DeclareRobustCommand{\VAN}[3]{##3}\VANthebibliography}

\usepackage{graphicx}	
\usepackage{amsmath}	
\usepackage{bm}

\title[Superorbital periods of Be/X-ray binaries]{Superorbital periods of Be/X-ray binaries driven by stellar spin precession}

\author[R. G. Martin ]{
Rebecca G. Martin\thanks{E-mail: rebecca.martin@unlv.edu}
\\
Nevada Center for Astrophysics, University of Nevada, Las Vegas,
4505 South Maryland Parkway, Las Vegas, NV 89154, USA\\
Department of Physics and Astronomy, University of Nevada, Las Vegas,
4505 South Maryland Parkway, Las Vegas, NV 89154, USA\\
}

\date{Accepted XXX. Received YYY; in original form ZZZ}

\pubyear{2023}

\begin{document}
\label{firstpage}
\pagerange{\pageref{firstpage}--\pageref{lastpage}}
\maketitle

\begin{abstract}
Superorbital periods are observed in the optical light curves of many Be/X-ray binaries yet their origin has remained somewhat elusive. We suggest that precession of the spin axis of the Be star can drive superorbital periods, particularly for short orbital period binaries.  We consider  the  short orbital period ($P_{\rm orb}=16.6\,\rm day$) and highly eccentric ($e_{\rm b}=0.72$) Be/X-ray binary A0538–66 that has a superorbital period of $421\,\rm day$. First we show that the spin axis precession timescale is about twice the observed superorbital period. Then, with hydrodynamic simulations we show that the Be star decretion disc can remain locked to the equator of the precessing Be star. At each  periastron passage of the neutron star, material is accreted into  a disc around the neutron star. The neutron star disc nodally precesses on the same timescale as the Be star disc and therefore both discs can contribute to the observed superorbital period.  For wider and less eccentric binary systems, the Be star disc can have a larger radial extent and more complex behaviour is expected as a result of disc warping and breaking.  
\end{abstract}




\section{Introduction}

Be/X-ray binaries consist of a Be type star in a binary most often with a neutron star companion \citep{Negueruela1998,Coe2005, Liu2005,haberl2016}. Be type stars are rapidly rotating  \citep{Slettebak1982,Porter1996,Porter2003}. They have a viscous Keplerian decretion disc \citep{Pringle1991} that forms from the ejection of material from the stellar equator \citep{Lee1991,Hanuschik1996,Carciofi2011}.  
The spin of the Be star is often misaligned to the binary orbital plane \citep[e.g.][]{Hughes1999,Hirata2007}. A small asymmetry in the supernova explosion that formed the neutron star leads to an eccentric binary orbit and a misaligned Be star spin \citep{Brandt1995,Martinetal2009b,Salvesen2020}. A disc can also form around the neutron star as it accretes material via Roche lobe overflow from the Be star disc \citep{Hayasaki2004, Franchini2021}.

Superorbital periods are observed in the optical light curves of many Be/X-ray binaries \citep[e.g.][]{Ogilvie2001,McGowan2008,Rajoelimanana2011}. The optical signature may be a combination of the Be star, the Be star disc and the neutron star disc \citep{Bird2012}. One of the most well studied objects, A0538-66, has an orbital period of $16.6\,\rm day$ and a stable superorbital period of $421\,\rm day$ \citep{Alcock2001,Rajoelimanana2017}. The origin of superorbital periods has been suggested to be the formation and depletion of the Be star disc \citep{McGowan2003}. However,  the correlation between the superorbital period and the orbital period \citep{Rajoelimanana2011,Townsend2020} suggests that there is a dynamical origin for the superorbital periods.

 A misaligned neutron star orbit provides a torque on the Be star disc that drives nodal precession about the binary angular momentum vector \citep{Lubow2000,Bateetal2000,Ogilvie2001b,martin2011be}. In the presence of dissipation within the disc, this leads to coplanar alignment of the disc to the binary orbital plane. While accretion is actively occurring from the Be star to the disc, the disc feels a torque from the addition of material to the disc that flows from the Be star equator. This torque aligns the disc to the Be star equator.  As a result of these competing torques, the disc may be subject to warping or breaking if it is not in sufficiently good radial communication \citep{Nixonetal2013,Dogan2015,Suffak2022}. A disc can precess like a solid body when the communication timescale is shorter than the precession timescale \citep{Papaloizou1983,PL1995,Papaloizou1995,Larwoodetal1996,Ogilvie1999}. Since Be star discs have a larger \cite{SS1973} $\alpha$ parameter than their disc aspect ratio, 
 the communication through the disc is likely through viscosity \citep{Nixon2016}.
 Hydrodynamic simulations of Be/X-ray binaries have shown that while accretion is actively occurring into the Be star disc, the disc does not undergo significant nodal precession with respect to the spin of the Be star once it has reached a steady state, unless it is radially extended enough to break \citep{Suffak2022}.

In this Letter, we point out that the spin axis of the tilted Be star undergoes precession about the binary angular momentum vector and this has been previously neglected in models of these systems. In Section~\ref{analytic} we show that the timescale for this precession may be similar to twice the observed superorbital period in the Be/X-ray binary A0538-66. In Section~\ref{simulation} we then perform a hydrodynamical simulation of the A0538-66 binary. The model includes the injection of material in to the disc at the Be star equator that precesses on a timescale of twice the superorbital period. The disc remains locked to the Be star equator and therefore we suggest that this may explain the superorbital periods, at least for the short orbital period binaries. Finally in Section~\ref{concs} we draw our conclusions and discuss the limitations of the model.

\section{Precession of the spin axis of the Be star}
\label{analytic}

We examine a binary consisting of a Be star of mass $M_1$ and radius $R_\star$ in an orbit with a neutron star of mass $M_2$ with binary semi-major axis $a_{\rm b}$ and eccentricity $e_{\rm b}$. The Be star spins with frequency $\Omega_\star=k_{\rm spin} \sqrt{G M_1/R_\star^3}$, where $k_{\rm spin}$ is a constant that is typically observed to be $>0.75$ \citep{Porter1996,Rivinius2013}. The spin axis is tilted to the binary orbital axis by inclination $i$ and precesses in a retrograde direction about the binary angular momentum vector on a timescale  
\begin{equation}
    t_{\rm prec}=\frac{2\pi}{\omega}.
\end{equation}
The precession frequency of the stellar spin axis about the binary angular momentum vector is
\begin{equation}
    \omega=\frac{3}{2} k \Omega_\star \left(\frac{M_2}{M_1}\right)\left(\frac{R_\star}{a_{\rm b} (1-e_{\rm b}^2)^{1/2}}\right)^3 \cos i
\end{equation}
\citep[e.g.][]{Lai1993,Lai1994}, where $k\approx 0.5$ is a constant that depends upon the stellar interior \citep[e.g.][]{Lai2014,Zanazzi2018}. 

 We  apply this to the Be/X-ray binary A0538–66 \citep{White1978,Johnston1979,Johnston1980,Skinner1980, Ducci2016,Ducci2022}. We take $M_1=8.84\,\rm M_\odot$ and $M_2=1.44\,\rm M_\odot$ \citep{Rajoelimanana2017}. The orbital period is $P_{\rm orb}=16.6\,\rm day$ \citep{Skinner1981} and the binary eccentricity is $e_{\rm b}=0.72$ \citep{Charles1983,Rajoelimanana2017}. With these parameters, the binary semi-major axis is $a_{\rm b}=59.6\,\rm R_\odot$ and we find 
\begin{align}
    t_{\rm prec}= &~ 890
    \left(\frac{M_1+M_2}{10.28\,\rm M_\odot} \right)
      \left(\frac{M_1}{8.84\,\rm M_\odot} \right)^{1/2}
       \left(\frac{M_2}{1.44\,\rm M_\odot} \right)^{-1}
        \left(\frac{k_{\rm spin}}{0.8} \right)^{-1} \notag \\ 
 & \times          \left(\frac{P_{\rm orb}}{16.6\,\rm day} \right)^{2}   
 \left( \frac{R_\star}{10\,\rm R_\odot}\right)^{-3/2}
 \left( \frac{1-e_{\rm b}^2}{0.48}\right)^{3/2}
 \left(\frac{k}{0.5}\right)^{-1}
    \,\rm day,
    \label{tsup}
\end{align}
where we have taken  $\cos i \approx 1$.
If the disc remains locked to the Be star equator, then the superorbital period is half of the precession timescale, $t_{\rm super}=t_{\rm prec}/2 = 445\,\rm day$, because the optical brightness of the system should be the same with the disc at longitude of ascending node $\Omega$ and at $(\Omega+180^\circ)$.
This is similar to the observed superorbital period of $421\,\rm day$ \citep{Rajoelimanana2017}. Some parameters are uncertain such as the radius of the Be star, the masses of the stars, the spin of the Be star, the inclination of the disc and the constant $k$. However, for reasonable parameters, the spin precession timescale may be in agreement with the observed superorbital period. 

We note that equation~(\ref{tsup}) suggests that the superorbital period scales with the orbital period squared. Observations show a large scatter but the best fitting line is more like superorbital period scaling with orbital period, at least for larger orbital periods \citep{Rajoelimanana2011,Townsend2020}. \cite{Bird2012} cautioned about the reliability of the measurements for the large orbital period systems. In this work, we consider only small orbital period binaries in which the disc is not radially extended. In the next section we consider a hydrodynamic model for such a system.

\section{Hydrodynamic simulation}
\label{simulation}

We model the evolution of a Be star disc in a binary with the smoothed particle hydrodynamics (SPH) code {\sc phantom} \citep{Price2010,Price2012a,Price2018} in order to examine the alignment of the Be star disc with the competing torques from material injection and the companion. The binary components are spherical sink particles that accrete the mass and angular momentum of any gas particles that pass inside of their sink radii \citep{Bateetal1995}. The accretion radius for the Be star is $8\,\rm R_\odot$ and for the neutron star is $1\,\rm R_\odot$. 
The gas particles each have a mass of $5\times 10^{-15}\,\rm M_\odot$. 
 Material is added to the disc with Keplerian velocity in a ring at an injection radius of $R_{\rm inj}=10\,\rm R_\odot$ at a rate of $\dot M_{\rm inj}=10^{-8}\,\rm M_\odot\,yr^{-1}$. The injection radius is chosen to be away from the Be star sink particle to increase the resolution of the simulation and to prevent implausibly large accretion rates back on to the Be star \citep{Nixon2020,Nixon2021}. Assuming that material is injected into a Be star disc at the surface of the star, the torque provided on the disc from the accretion in this simulation is equivalent to that from a star with a radius of $10\,\rm R_\odot$. Inside of the injection radius the disc is an accretion disc, while outside of the injection radius it behaves as a decretion disc \citep[e.g.][]{ML2011,Rimulo2018}. For comparison, we also considered a simulation with a smaller injection radius of $R_{\rm inj}=8.3\,\rm R_\odot$ \citep[in line with previous simulations that include injection of particles into the Be star disc, e.g.][]{Cyr2017,Suffak2022}. We found that the smaller injection radius does not change our conclusions on the disc alignment. However,  the number of particles in the disc was reduced by a factor of about 8. For a wider orbit binary, where warping may be possible, the injection radius could have a larger influence as discussed in Section~\ref{concs}. 

\begin{figure}
\begin{center}
\includegraphics[width=0.9\columnwidth]{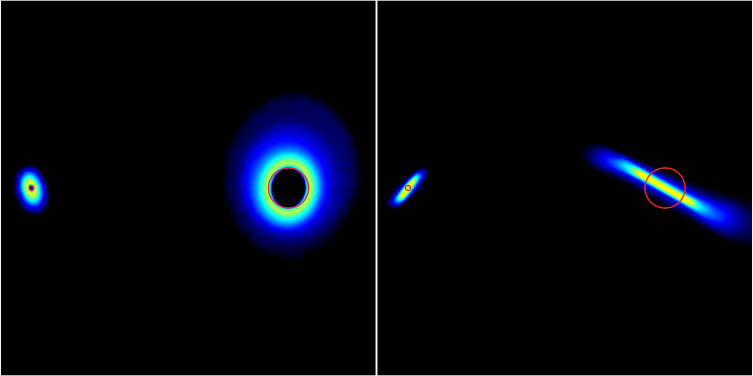}
\includegraphics[width=0.9\columnwidth]{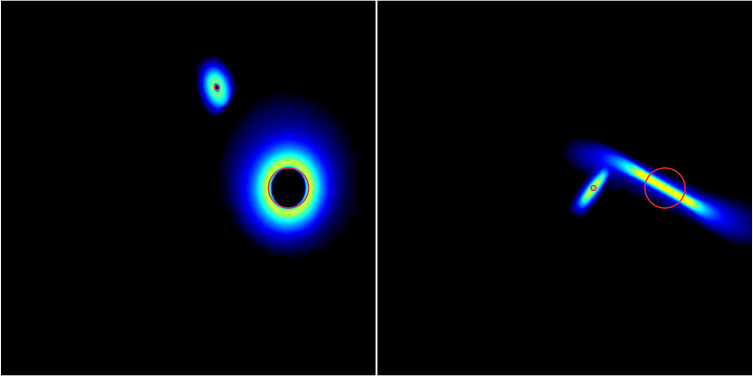}
\includegraphics[width=0.9\columnwidth]{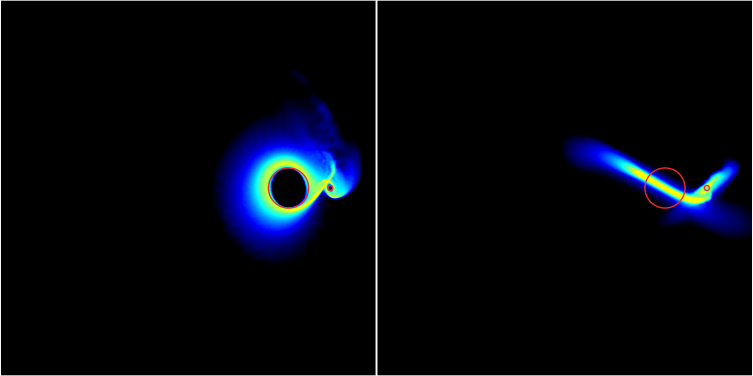}
\includegraphics[width=0.9\columnwidth]{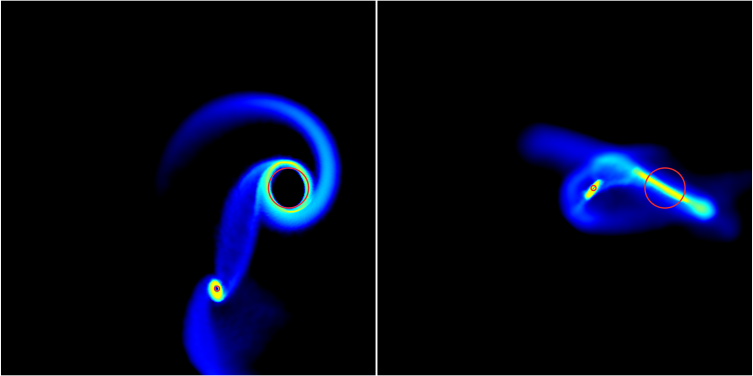}
\includegraphics[width=0.9\columnwidth]{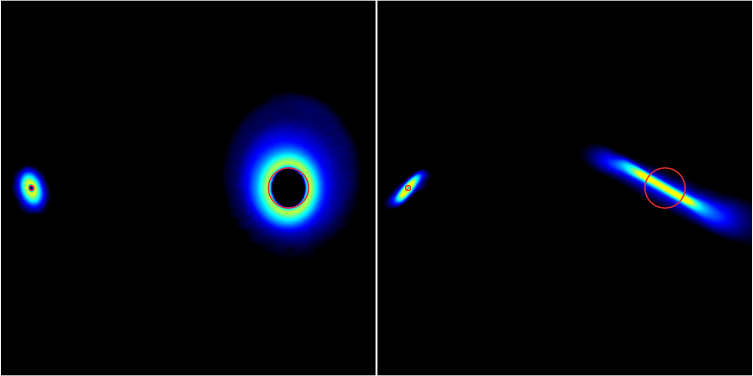}
	\end{center}
    \caption{The disc evolution over one binary orbital period. The red open circles show the outline of the spherical Be star (larger circle) and the neutron star (smaller circle) with the size scaled to their accretion radius. The gas colour shows the column density with yellow being about two orders of magnitude higher than blue. In each panel, the left hand side shows the $x-y$ plane in which the binary orbits while the right hand side shows the $x-z$ plane. From top to bottom, the times shown are $t=12.0$ (apastron), $12.4$, $12.5$ (periastron), $12.6$ and $13.0\,P_{\rm orb}$ (apastron).
    }
    \label{splashorb}
\end{figure}

\begin{figure}
\begin{center}
\includegraphics[width=1\columnwidth]{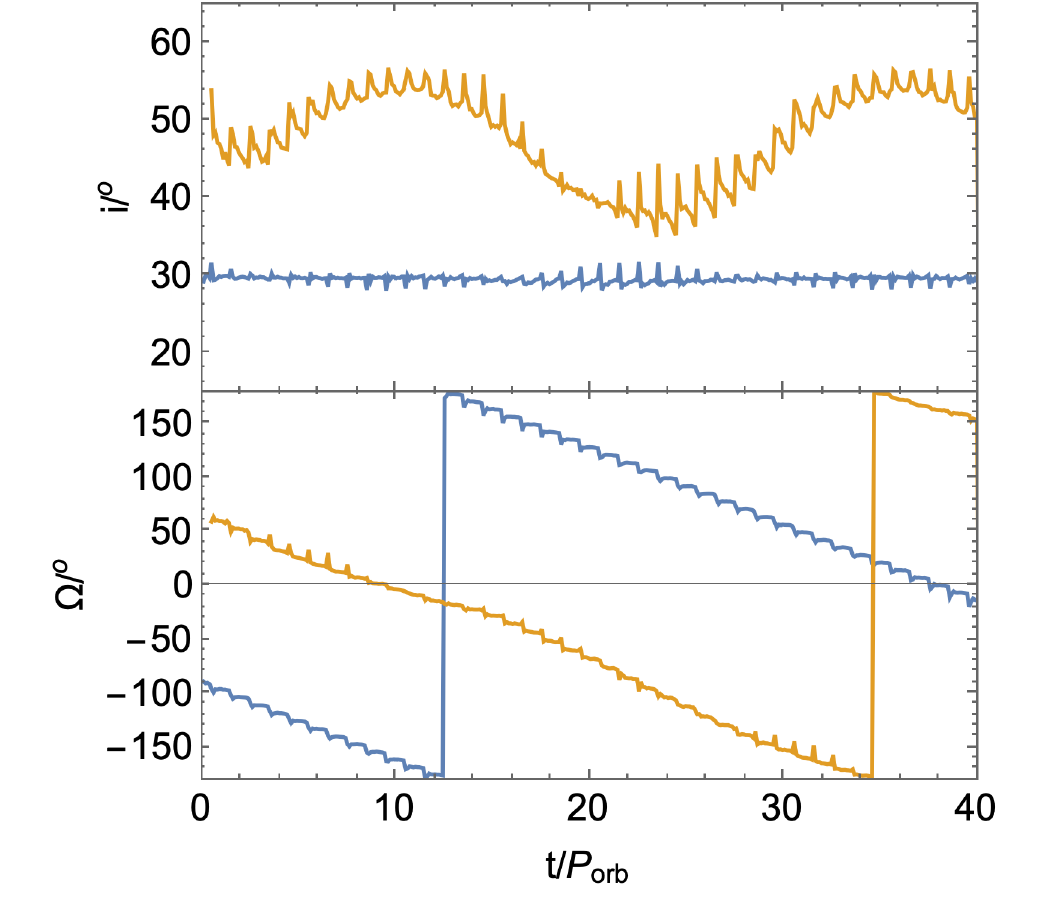}
	\end{center}
    \caption{The inclination (upper panel) and longitude of ascending node (lower panel) of the Be star disc (blue) and the neutron star disc (orange). 
    }
    \label{sigma}
\end{figure}

The spin axis precession of the Be star is modelled through the injection of material  in to the Be star disc  from  the preccesing equator of the Be star. The particles are injected with Keplerian velocity around the instantaneous Be star spin axis.   The spin axis of the Be star is tilted by $i=30^\circ$ to the binary orbit and it nodally precesses in a retrograde direction with a timescale $t_{\rm prec}=2\, t_{\rm super}=842\,{\rm day}=50.7\, P_{\rm orb}$.
The binary orbit is in the $x-y$ plane. Particles are added with a random azimuthal angle, $\phi$, in the range $0-2\pi$ around a ring at radius $R_{\rm inj}$ around the Be star. In order to find the injection position and location for each particle, we consider a ring in the $x-y$ plane that is tilted in the $x-z$ plane by $i=30^\circ$  and then rotated about the $z$ axis by angle $\beta=-2 \pi t/t_{\rm prec}$. 
Therefore, the position of an injected particle relative to the Be star is
     \begin{align}
    \bm{r} &= R_{\rm inj} \begin{pmatrix}
           \cos \beta \cos \phi \cos i-\sin \beta \sin  \phi \\
           \sin \beta \cos \phi \cos i+\cos \beta \sin \phi \\
           \cos \phi \sin i 
         \end{pmatrix}
  \end{align}
and its velocity is
   \begin{align}
    \bm{v} &= v_{\rm K}\begin{pmatrix}
           -\cos \beta \sin \phi \cos i-\sin \beta \cos  \phi \\
          - \sin \beta \sin \phi \cos i+\cos \beta \cos \phi \\
           -\sin \phi \sin i 
         \end{pmatrix},
  \end{align}
  where the Keplerian velocity around the Be star is $v_{\rm K}=\sqrt{\frac{G M_1}{R_{\rm inj}}} $.

 In the code, the viscosity is  implemented by adapting the SPH artificial viscosity \citep{Lodato2010} with $\alpha_{\rm AV}=1.0$ and $\beta_{\rm AV}=2.0$.  The disc is isothermal with $H/R=0.04$ at $R=8\,\rm R_\odot$. The binary begins at apastron and orbits in the $x-y$ plane. Since the disc mass is very small compared to the binary mass, the binary orbit does not change during the simulation. The disc mass oscillates over the orbital period.  Once a quasi-steady-state has been reached, at apastron the  mass of the Be star disc is approximately $4\times 10^{-10}\,\rm M_\odot$ and the mass of the neutron star disc is about $10^{-10}\,\rm M_\odot$. The density averaged smoothing length in the Be star disc once the disc has reached a steady state is $\left<h\right>/H\approx 0.5$ which is equivalent to a \cite{SS1973} $\alpha \approx 0.05$ \citep[see equation 11 in][]{Okazaki2002}. Observationally the viscosity of a Be star disc is large, $\alpha \approx 0.1-0.3$ \citep{Jones2008,Rimulo2018,Ghoreyshi2018,Martin2019,Granada2021}. Once the system has reached a steady state, about $80\%$ of the injected mass is accreted on to the Be star, $10\%$ is accreted on to the neutron star and about $10\%$ forms very low density circumbinary material \citep[this has been seen before, e.g.][]{Franchini2019}. We note that the resolution of the neutron star disc is  $\left<h\right>/H\approx 1.0$, and exploring its evolution in detail would require much higher resolution (computationally expensive) simulations \citep[see also discussions in][]{Martinetal2014, Franchini2021}.  

Fig.~\ref{splashorb} shows the disc evolution over one orbital period from a time  of $t=12.0\,P_{\rm orb}$ up to $t=13.0\,P_{\rm orb}$. In the upper panels the binary is at apastron, in the middle panels it is at periastron and in the lower panels it is again at apastron.  Since the periastron separation is small ($16.7\,\rm R_\odot$), the Be star disc cannot be very radially extended \citep[e.g.][]{Reig1997,Okazaki2001}. The neutron star accretes material from the Be star at each periastron passage, as seen by the streams between the two stars in the fourth row. A disc forms around the neutron star that is also misaligned to the binary orbit. The orientations of both discs  do not change significantly over one orbital period.

Fig.~\ref{sigma} shows the evolution of the inclination and longitude of ascending node of both discs. The Be star disc properties are calculated at a radius of $R=12\,\rm R_\odot$ from the Be star while the neutron star disc properties are calculated at $R=4\,\rm R_\odot$ from the neutron star. Both discs undergo small oscillations in inclination and longitude of ascending node on the orbital period. The inclination of the Be star disc remains roughly constant because of the injection of material at the inclination of $i=30^\circ$. The disc around the neutron star forms at a higher inclination to the Be star disc. This has been seen before in simulations \citep[e.g.][]{Franchini2019}.  The neutron star disc inclination oscillates on half the disc precession timescale, which is equal to the superorbital period. The disc around the neutron star also undergoes nodal precession on exactly the same timescale as the Be star disc since it forms from accretion of material from the Be star disc.  Therefore contributions to the observed optical superorbital periods could be coming from both discs since the changes are on the same timescale.  

We do not see any evidence of von Zeipel-Kozai-Lidov \citep[ZKL,][]{vonZeipel1910,Kozai1962,Lidov1962} oscillations in the neutron star disc  despite the high inclination disc around the neutron star.  This is because the material flows through the disc on a timescale shorter than the ZKL oscillation timescale which could be a result of low resolution in that disc \citep[see also discussion in][]{Smallwood2023}. ZKL oscillations of a highly inclined fluid disc involve exchange of global disc inclination and eccentricity \citep{Martinetal2014b,Fu2015,Lubow2017,Zanazzi2017}.

\begin{figure}
\begin{center}
\includegraphics[width=0.9\columnwidth]{splash130.png}
\includegraphics[width=0.9\columnwidth]{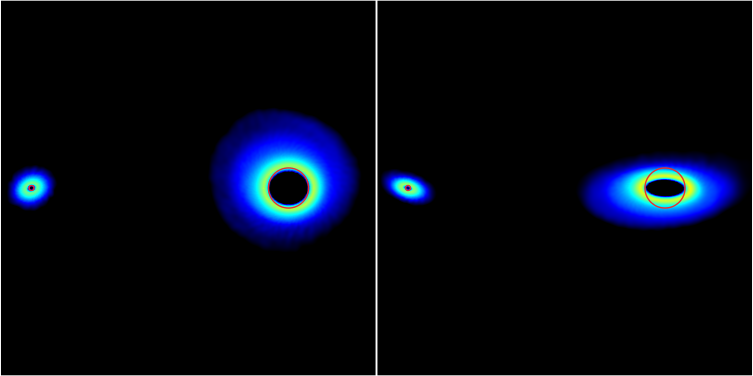}
\includegraphics[width=0.9\columnwidth]{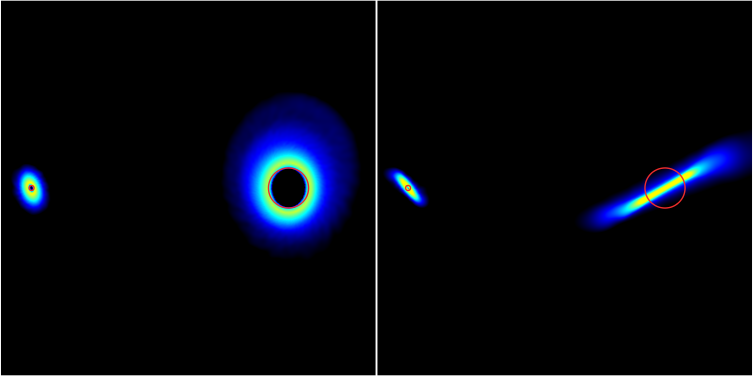}
	\end{center}
    \caption{The disc evolution over one superorbital period. Same as Fig.~\ref{splashorb} except the times shown from top to bottom are $t=13$, $25$ and $38\,P_{\rm orb}$ (all times shown are at apastron separation of the binary). The precession timescale for the stellar spin axis is $t_{\rm prec}=50.7\, P_{\rm orb}$. This produces a superorbital period of $t_{\rm super}=25.3\, P_{\rm orb}$.
    }
    \label{splashsuperorb}
\end{figure}

Fig.~\ref{splashsuperorb} shows the disc evolution over one superorbital period. In the right hand panels, the disc is initially seen edge on. From this viewpoint, this is the minimum brightness in the optical light curve.  In the middle right panels, the Be star spin is pointing towards the observer and the disc appears larger and therefore brighter. This corresponds to the optical peak. In the lower panels the Be star spin has precessed by $180^\circ$ from the top panel and the disc is again seen edge on and the brightness is at the minimum again. 

There is very little warp or twist in either disc except in the outer parts of the Be star disc when then neutron star is close to periastron. Because of the small radial extent of the Be star disc in the short orbital period and eccentric binary, the  mass accretion torque dominates the disc dynamics and the disc remains locked to the equator of the Be star. For a wider binary, more complex behaviour is expected since a radially extended disc can undergo warping and breaking \citep[e.g.][]{martin2011be,NK2012}. In this case, the precession timescale for the disc is not so easily linked to the spin precession timescale and depends upon the tidal torque and the level of warping.

\section{Conclusions}
\label{concs}

A rapidly rotating Be star ejects material from its equator into a decretion disc. If the spin axis of the Be star is misaligned to the orbital plane of a companion, then a misaligned disc can form. A misaligned Be star disc  feels two competing torques. First, there is the torque from the accretion of new material from the Be star to the disc. This acts to align the disc to the Be star equator. This torque is sensitive to the injection radius of the material. Secondly, the torque from the neutron star causes nodal precession of the disc about the binary angular momentum vector which leads to coplanar alignment of the disc to the binary orbital plane in the presence of dissipation in the disc.   While a Be star decretion disc is actively being fed material from the equator of the Be star, it can remain locked to the equator of the Be star if the disc is radially narrow and the mass accretion torque dominates the tidal torque. This occurs for systems with a small binary periastron separation.

The nodal precession timescale of a disc that is locked to the Be star equator is therefore equal to the timescale for spin axis precession of the Be star.  The spin axis precession has been previously neglected in the literature but it is especially important for short orbital period binaries. We show that the axial precession timescale for the Be star may be similar to the twice the observed superorbital period in Be/X-ray binaries. Therefore we suggest that superorbital periods in short orbital period systems are driven by the precession of a tilted disc that precesses with the spin axis of the Be star.  Moreover, the disc that forms around the neutron star precesses on the same timescale since its dynamics are dominated by the accretion of material at periastron separation. Therefore both discs can contribute to the observed superorbital period.

In a Be/X-ray binary with a longer orbital period, or a more circular orbit than considered here, the disc dynamics could be quite different. A radially extended disc may not remain locked to the equator and could become warped or even broken \citep{Suffak2022}. The outer parts of a warped disc, or the outer ring of a broken disc, could precess on the timescale induced by the tidal torque of the companion \citep[e.g.][]{martin2011be,marr2022,martin2022pleione} while the precession timescale for the inner parts could be dominated by the stellar spin axis precession. Finally we note that superorbital periods driven by this mechanism depend upon the viewing inclination of the system. If the binary orbit is close to face on, then superorbital periods will not be seen unless the spin axis tilt is very large.

\section*{Acknowledgements}
RGM thanks Phil Charles, David Vallet and Stephen Lepp for useful conversations and an anonymous referee for useful comments.  Computer support was provided by UNLV’s National Supercomputing Center. We acknowledge support from NASA through grant 80NSSC21K0395. We acknowledge the use of SPLASH \citep{Price2007} for the rendering of Figs.~\ref{splashorb} and~\ref{splashsuperorb}.  

\section*{Data Availability}

 The data underlying this article will be shared on reasonable request to the corresponding author.
 



\bibliographystyle{mnras}
\bibliography{martin} 

\begin{thebibliography}{}
\makeatletter
\relax
\def\mn@urlcharsother{\let\do\@makeother \do\$\do\&\do\#\do\^\do\_\do\%\do\~}
\def\mn@doi{\begingroup\mn@urlcharsother \@ifnextchar [ {\mn@doi@}
  {\mn@doi@[]}}
\def\mn@doi@[#1]#2{\def\@tempa{#1}\ifx\@tempa\@empty \href
  {http://dx.doi.org/#2} {doi:#2}\else \href {http://dx.doi.org/#2} {#1}\fi
  \endgroup}
\def\mn@eprint#1#2{\mn@eprint@#1:#2::\@nil}
\def\mn@eprint@arXiv#1{\href {http://arxiv.org/abs/#1} {{\tt arXiv:#1}}}
\def\mn@eprint@dblp#1{\href {http://dblp.uni-trier.de/rec/bibtex/#1.xml}
  {dblp:#1}}
\def\mn@eprint@#1:#2:#3:#4\@nil{\def\@tempa {#1}\def\@tempb {#2}\def\@tempc
  {#3}\ifx \@tempc \@empty \let \@tempc \@tempb \let \@tempb \@tempa \fi \ifx
  \@tempb \@empty \def\@tempb {arXiv}\fi \@ifundefined
  {mn@eprint@\@tempb}{\@tempb:\@tempc}{\expandafter \expandafter \csname
  mn@eprint@\@tempb\endcsname \expandafter{\@tempc}}}

\bibitem[\protect\citeauthoryear{{Alcock} et~al.,}{{Alcock}
  et~al.}{2001}]{Alcock2001}
{Alcock} C.,  et~al., 2001, \mn@doi [\mnras]
  {10.1046/j.1365-8711.2001.04041.x}, \href
  {https://ui.adsabs.harvard.edu/abs/2001MNRAS.321..678A} {321, 678}

\bibitem[\protect\citeauthoryear{{Bate}, {Bonnell}  \& {Price}}{{Bate}
  et~al.}{1995}]{Bateetal1995}
{Bate} M.~R.,  {Bonnell} I.~A.,   {Price} N.~M.,  1995, \mn@doi [\mnras]
  {10.1093/mnras/277.2.362}, \href
  {http://adsabs.harvard.edu/abs/1995MNRAS.277..362B} {277, 362}

\bibitem[\protect\citeauthoryear{{Bate}, {Bonnell}, {Clarke}, {Lubow},
  {Ogilvie}, {Pringle}  \& {Tout}}{{Bate} et~al.}{2000}]{Bateetal2000}
{Bate} M.~R.,  {Bonnell} I.~A.,  {Clarke} C.~J.,  {Lubow} S.~H.,  {Ogilvie}
  G.~I.,  {Pringle} J.~E.,   {Tout} C.~A.,  2000, \mn@doi [MNRAS]
  {10.1046/j.1365-8711.2000.03648.x}, \href
  {http://adsabs.harvard.edu/abs/2000MNRAS.317..773B} {317, 773}

\bibitem[\protect\citeauthoryear{{Bird}, {Coe}, {McBride}  \& {Udalski}}{{Bird}
  et~al.}{2012}]{Bird2012}
{Bird} A.~J.,  {Coe} M.~J.,  {McBride} V.~A.,   {Udalski} A.,  2012, \mn@doi
  [\mnras] {10.1111/j.1365-2966.2012.21163.x}, \href
  {https://ui.adsabs.harvard.edu/abs/2012MNRAS.423.3663B} {423, 3663}

\bibitem[\protect\citeauthoryear{{Brandt} \& {Podsiadlowski}}{{Brandt} \&
  {Podsiadlowski}}{1995}]{Brandt1995}
{Brandt} N.,  {Podsiadlowski} P.,  1995, \mn@doi [\mnras]
  {10.1093/mnras/274.2.461}, \href
  {http://adsabs.harvard.edu/abs/1995MNRAS.274..461B} {274, 461}

\bibitem[\protect\citeauthoryear{{Carciofi}}{{Carciofi}}{2011}]{Carciofi2011}
{Carciofi} A.~C.,  2011, in {Neiner} C.,  {Wade} G.,  {Meynet} G.,   {Peters}
  G.,  eds,  IAU Symposium Vol. 272, IAU Symposium. pp 325--336 (\mn@eprint
  {arXiv} {1009.3969}), \mn@doi{10.1017/S1743921311010738}

\bibitem[\protect\citeauthoryear{{Charles} et~al.,}{{Charles}
  et~al.}{1983}]{Charles1983}
{Charles} P.~A.,  et~al., 1983, \mn@doi [\mnras] {10.1093/mnras/202.3.657},
  \href {https://ui.adsabs.harvard.edu/abs/1983MNRAS.202..657C} {202, 657}

\bibitem[\protect\citeauthoryear{{Coe}, {Edge}, {Galache}  \& {McBride}}{{Coe}
  et~al.}{2005}]{Coe2005}
{Coe} M.~J.,  {Edge} W.~R.~T.,  {Galache} J.~L.,   {McBride} V.~A.,  2005,
  \mn@doi [\mnras] {10.1111/j.1365-2966.2004.08467.x}, \href
  {https://ui.adsabs.harvard.edu/abs/2005MNRAS.356..502C} {356, 502}

\bibitem[\protect\citeauthoryear{{Cyr}, {Jones}, {Panoglou}, {Carciofi}  \&
  {Okazaki}}{{Cyr} et~al.}{2017}]{Cyr2017}
{Cyr} I.~H.,  {Jones} C.~E.,  {Panoglou} D.,  {Carciofi} A.~C.,   {Okazaki}
  A.~T.,  2017, \mn@doi [\mnras] {10.1093/mnras/stx1427}, \href
  {https://ui.adsabs.harvard.edu/abs/2017MNRAS.471..596C} {471, 596}

\bibitem[\protect\citeauthoryear{{Do{\u{g}}an}, {Nixon}, {King}  \&
  {Price}}{{Do{\u{g}}an} et~al.}{2015}]{Dogan2015}
{Do{\u{g}}an} S.,  {Nixon} C.,  {King} A.,   {Price} D.~J.,  2015, \mn@doi
  [\mnras] {10.1093/mnras/stv347}, \href
  {https://ui.adsabs.harvard.edu/abs/2015MNRAS.449.1251D} {449, 1251}

\bibitem[\protect\citeauthoryear{{Ducci}, {Covino}, {Doroshenko}, {Mereghetti},
  {Santangelo}  \& {Sasaki}}{{Ducci} et~al.}{2016}]{Ducci2016}
{Ducci} L.,  {Covino} S.,  {Doroshenko} V.,  {Mereghetti} S.,  {Santangelo} A.,
    {Sasaki} M.,  2016, \mn@doi [\aap] {10.1051/0004-6361/201629236}, \href
  {https://ui.adsabs.harvard.edu/abs/2016A&A...595A.103D} {595, A103}

\bibitem[\protect\citeauthoryear{{Ducci} et~al.,}{{Ducci}
  et~al.}{2022}]{Ducci2022}
{Ducci} L.,  et~al., 2022, \mn@doi [\aap] {10.1051/0004-6361/202140867}, \href
  {https://ui.adsabs.harvard.edu/abs/2022A&A...661A..22D} {661, A22}

\bibitem[\protect\citeauthoryear{{Franchini} \& {Martin}}{{Franchini} \&
  {Martin}}{2021}]{Franchini2021}
{Franchini} A.,  {Martin} R.~G.,  2021, \mn@doi [\apjl]
  {10.3847/2041-8213/ac4029}, \href
  {https://ui.adsabs.harvard.edu/abs/2021ApJ...923L..18F} {923, L18}

\bibitem[\protect\citeauthoryear{{Franchini}, {Martin}  \& {Lubow}}{{Franchini}
  et~al.}{2019}]{Franchini2019}
{Franchini} A.,  {Martin} R.~G.,   {Lubow} S.~H.,  2019, \mn@doi [\mnras]
  {10.1093/mnras/stz424}, \href
  {https://ui.adsabs.harvard.edu/abs/2019MNRAS.485..315F} {485, 315}

\bibitem[\protect\citeauthoryear{{Fu}, {Lubow}  \& {Martin}}{{Fu}
  et~al.}{2015}]{Fu2015}
{Fu} W.,  {Lubow} S.~H.,   {Martin} R.~G.,  2015, \mn@doi [\apj]
  {10.1088/0004-637X/807/1/75}, \href
  {http://adsabs.harvard.edu/abs/2015ApJ...807...75F} {807, 75}

\bibitem[\protect\citeauthoryear{{Ghoreyshi} et~al.,}{{Ghoreyshi}
  et~al.}{2018}]{Ghoreyshi2018}
{Ghoreyshi} M.~R.,  et~al., 2018, \mn@doi [\mnras] {10.1093/mnras/sty1577},
  \href {https://ui.adsabs.harvard.edu/abs/2018MNRAS.479.2214G} {479, 2214}

\bibitem[\protect\citeauthoryear{{Granada}, {Jones}  \& {Sigut}}{{Granada}
  et~al.}{2021}]{Granada2021}
{Granada} A.,  {Jones} C.~E.,   {Sigut} T.~A.~A.,  2021, \mn@doi [\apj]
  {10.3847/1538-4357/ac222f}, \href
  {https://ui.adsabs.harvard.edu/abs/2021ApJ...922..148G} {922, 148}

\bibitem[\protect\citeauthoryear{{Haberl} \& {Sturm}}{{Haberl} \&
  {Sturm}}{2016}]{haberl2016}
{Haberl} F.,  {Sturm} R.,  2016, \mn@doi [\aap] {10.1051/0004-6361/201527326},
  \href {https://ui.adsabs.harvard.edu/abs/2016A&A...586A..81H} {586, A81}

\bibitem[\protect\citeauthoryear{{Hanuschik}}{{Hanuschik}}{1996}]{Hanuschik1996}
{Hanuschik} R.~W.,  1996, A\&A, \href
  {http://adsabs.harvard.edu/abs/1996A%26A...308..170H} {308, 170}

\bibitem[\protect\citeauthoryear{{Hayasaki} \& {Okazaki}}{{Hayasaki} \&
  {Okazaki}}{2004}]{Hayasaki2004}
{Hayasaki} K.,  {Okazaki} A.~T.,  2004, \mn@doi [\mnras]
  {10.1111/j.1365-2966.2004.07702.x}, \href
  {http://adsabs.harvard.edu/abs/2004MNRAS.350..971H} {350, 971}

\bibitem[\protect\citeauthoryear{{Hirata}}{{Hirata}}{2007}]{Hirata2007}
{Hirata} R.,  2007, in {Okazaki} A.~T.,  {Owocki} S.~P.,   {Stefl} S.,  eds,
  Astronomical Society of the Pacific Conference Series Vol. 361, Active
  OB-Stars: Laboratories for Stellare and Circumstellar Physics. p.~267

\bibitem[\protect\citeauthoryear{{Hughes} \& {Bailes}}{{Hughes} \&
  {Bailes}}{1999}]{Hughes1999}
{Hughes} A.,  {Bailes} M.,  1999, \mn@doi [\apj] {10.1086/307605}, \href
  {https://ui.adsabs.harvard.edu/abs/1999ApJ...522..504H} {522, 504}

\bibitem[\protect\citeauthoryear{{Johnston}, {Bradt}, {Doxsey}, {Griffiths},
  {Schwartz}  \& {Schwarz}}{{Johnston} et~al.}{1979}]{Johnston1979}
{Johnston} M.~D.,  {Bradt} H.~V.,  {Doxsey} R.~E.,  {Griffiths} R.~E.,
  {Schwartz} D.~A.,   {Schwarz} J.,  1979, \mn@doi [\apjl] {10.1086/182952},
  \href {https://ui.adsabs.harvard.edu/abs/1979ApJ...230L..11J} {230, L11}

\bibitem[\protect\citeauthoryear{{Johnston}, {Griffiths}  \& {Ward}}{{Johnston}
  et~al.}{1980}]{Johnston1980}
{Johnston} M.~D.,  {Griffiths} R.~E.,   {Ward} M.~J.,  1980, \mn@doi [\nat]
  {10.1038/285026a0}, \href
  {https://ui.adsabs.harvard.edu/abs/1980Natur.285...26J} {285, 26}

\bibitem[\protect\citeauthoryear{{Jones}, {Sigut}  \& {Porter}}{{Jones}
  et~al.}{2008}]{Jones2008}
{Jones} C.~E.,  {Sigut} T.~A.~A.,   {Porter} J.~M.,  2008, \mn@doi [\mnras]
  {10.1111/j.1365-2966.2008.13206.x}, \href
  {http://adsabs.harvard.edu/abs/2008MNRAS.386.1922J} {386, 1922}

\bibitem[\protect\citeauthoryear{{Kozai}}{{Kozai}}{1962}]{Kozai1962}
{Kozai} Y.,  1962, \mn@doi [AJ] {10.1086/108790}, \href
  {http://adsabs.harvard.edu/abs/1962AJ.....67..591K} {67, 591}

\bibitem[\protect\citeauthoryear{{Lai}}{{Lai}}{2014}]{Lai2014}
{Lai} D.,  2014, \mn@doi [MNRAS] {10.1093/mnras/stu485}, \href
  {http://adsabs.harvard.edu/abs/2014MNRAS.440.3532L} {440, 3532}

\bibitem[\protect\citeauthoryear{{Lai}, {Rasio}  \& {Shapiro}}{{Lai}
  et~al.}{1993}]{Lai1993}
{Lai} D.,  {Rasio} F.~A.,   {Shapiro} S.~L.,  1993, \mn@doi [\apjs]
  {10.1086/191822}, \href
  {https://ui.adsabs.harvard.edu/abs/1993ApJS...88..205L} {88, 205}

\bibitem[\protect\citeauthoryear{{Lai}, {Rasio}  \& {Shapiro}}{{Lai}
  et~al.}{1994}]{Lai1994}
{Lai} D.,  {Rasio} F.~A.,   {Shapiro} S.~L.,  1994, \mn@doi [\apj]
  {10.1086/173606}, \href
  {https://ui.adsabs.harvard.edu/abs/1994ApJ...420..811L} {420, 811}

\bibitem[\protect\citeauthoryear{{Larwood}, {Nelson}, {Papaloizou}  \&
  {Terquem}}{{Larwood} et~al.}{1996}]{Larwoodetal1996}
{Larwood} J.~D.,  {Nelson} R.~P.,  {Papaloizou} J.~C.~B.,   {Terquem} C.,
  1996, MNRAS, \href {http://adsabs.harvard.edu/abs/1996MNRAS.282..597L} {282,
  597}

\bibitem[\protect\citeauthoryear{{Lee}, {Osaki}  \& {Saio}}{{Lee}
  et~al.}{1991}]{Lee1991}
{Lee} U.,  {Osaki} Y.,   {Saio} H.,  1991, MNRAS, \href
  {http://adsabs.harvard.edu/abs/1991MNRAS.250..432L} {250, 432}

\bibitem[\protect\citeauthoryear{{Lidov}}{{Lidov}}{1962}]{Lidov1962}
{Lidov} M.~L.,  1962, \mn@doi [Planet. Space Sci.]
  {10.1016/0032-0633(62)90129-0}, \href
  {http://adsabs.harvard.edu/abs/1962P%26SS....9..719L} {9, 719}

\bibitem[\protect\citeauthoryear{{Liu}, {van Paradijs}  \& {van den
  Heuvel}}{{Liu} et~al.}{2005}]{Liu2005}
{Liu} Q.~Z.,  {van Paradijs} J.,   {van den Heuvel} E.~P.~J.,  2005, \mn@doi
  [\aap] {10.1051/0004-6361:20053718}, \href
  {https://ui.adsabs.harvard.edu/abs/2005A&A...442.1135L} {442, 1135}

\bibitem[\protect\citeauthoryear{{Lodato} \& {Price}}{{Lodato} \&
  {Price}}{2010}]{Lodato2010}
{Lodato} G.,  {Price} D.~J.,  2010, \mn@doi [\mnras]
  {10.1111/j.1365-2966.2010.16526.x}, \href
  {https://ui.adsabs.harvard.edu/abs/2010MNRAS.405.1212L} {405, 1212}

\bibitem[\protect\citeauthoryear{{Lubow} \& {Ogilvie}}{{Lubow} \&
  {Ogilvie}}{2000}]{Lubow2000}
{Lubow} S.~H.,  {Ogilvie} G.~I.,  2000, \mn@doi [\apj] {10.1086/309101}, \href
  {http://adsabs.harvard.edu/abs/2000ApJ...538..326L} {538, 326}

\bibitem[\protect\citeauthoryear{{Lubow} \& {Ogilvie}}{{Lubow} \&
  {Ogilvie}}{2017}]{Lubow2017}
{Lubow} S.~H.,  {Ogilvie} G.~I.,  2017, \mn@doi [\mnras]
  {10.1093/mnras/stx990}, \href
  {http://adsabs.harvard.edu/abs/2017MNRAS.469.4292L} {469, 4292}

\bibitem[\protect\citeauthoryear{{Marr}, {Jones}, {Tycner}, {Carciofi}  \&
  {Silva}}{{Marr} et~al.}{2022}]{marr2022}
{Marr} K.~C.,  {Jones} C.~E.,  {Tycner} C.,  {Carciofi} A.~C.,   {Silva}
  A.~C.~F.,  2022, \mn@doi [\apj] {10.3847/1538-4357/ac551b}, \href
  {https://ui.adsabs.harvard.edu/abs/2022ApJ...928..145M} {928, 145}

\bibitem[\protect\citeauthoryear{{Martin} \& {Lepp}}{{Martin} \&
  {Lepp}}{2022}]{martin2022pleione}
{Martin} R.~G.,  {Lepp} S.,  2022, \mn@doi [\mnras] {10.1093/mnrasl/slac090},
  \href {https://ui.adsabs.harvard.edu/abs/2022MNRAS.516L..86M} {516, L86}

\bibitem[\protect\citeauthoryear{{Martin} \& {Lubow}}{{Martin} \&
  {Lubow}}{2011}]{ML2011}
{Martin} R.~G.,  {Lubow} S.~H.,  2011, \mn@doi [\mnras]
  {10.1111/j.1365-2966.2011.18228.x}, \href
  {https://ui.adsabs.harvard.edu/abs/2011MNRAS.413.1447M} {413, 1447}

\bibitem[\protect\citeauthoryear{{Martin}, {Tout}  \& {Pringle}}{{Martin}
  et~al.}{2009}]{Martinetal2009b}
{Martin} R.~G.,  {Tout} C.~A.,   {Pringle} J.~E.,  2009, \mn@doi [MNRAS]
  {10.1111/j.1365-2966.2009.15031.x}, \href
  {http://adsabs.harvard.edu/abs/2009MNRAS.397.1563M} {397, 1563}

\bibitem[\protect\citeauthoryear{{Martin}, {Pringle}, {Tout}  \&
  {Lubow}}{{Martin} et~al.}{2011}]{martin2011be}
{Martin} R.~G.,  {Pringle} J.~E.,  {Tout} C.~A.,   {Lubow} S.~H.,  2011,
  \mn@doi [\mnras] {10.1111/j.1365-2966.2011.19231.x}, \href
  {https://ui.adsabs.harvard.edu/abs/2011MNRAS.416.2827M} {416, 2827}

\bibitem[\protect\citeauthoryear{{Martin}, {Nixon}, {Armitage}, {Lubow}  \&
  {Price}}{{Martin} et~al.}{2014a}]{Martinetal2014}
{Martin} R.~G.,  {Nixon} C.,  {Armitage} P.~J.,  {Lubow} S.~H.,   {Price}
  D.~J.,  2014a, \mn@doi [ApJL] {10.1088/2041-8205/790/2/L34}, \href
  {http://adsabs.harvard.edu/abs/2014ApJ...790L..34M} {790, L34}

\bibitem[\protect\citeauthoryear{{Martin}, {Nixon}, {Lubow}, {Armitage},
  {Price}, {Do{\u g}an}  \& {King}}{{Martin} et~al.}{2014b}]{Martinetal2014b}
{Martin} R.~G.,  {Nixon} C.,  {Lubow} S.~H.,  {Armitage} P.~J.,  {Price} D.~J.,
   {Do{\u g}an} S.,   {King} A.,  2014b, \mn@doi [ApJL]
  {10.1088/2041-8205/792/2/L33}, \href
  {http://adsabs.harvard.edu/abs/2014ApJ...792L..33M} {792, L33}

\bibitem[\protect\citeauthoryear{{Martin}, {Nixon}, {Pringle}  \&
  {Livio}}{{Martin} et~al.}{2019}]{Martin2019}
{Martin} R.~G.,  {Nixon} C.~J.,  {Pringle} J.~E.,   {Livio} M.,  2019, \mn@doi
  [New Astronomy] {10.1016/j.newast.2019.01.001}, \href
  {http://adsabs.harvard.edu/abs/2019NewA...70....7M} {70, 7}

\bibitem[\protect\citeauthoryear{{McGowan} \& {Charles}}{{McGowan} \&
  {Charles}}{2003}]{McGowan2003}
{McGowan} K.~E.,  {Charles} P.~A.,  2003, \mn@doi [\mnras]
  {10.1046/j.1365-8711.2003.06220.x}, \href
  {https://ui.adsabs.harvard.edu/abs/2003MNRAS.339..748M} {339, 748}

\bibitem[\protect\citeauthoryear{{McGowan}, {Coe}, {Schurch}, {Corbet},
  {Galache}  \& {Udalski}}{{McGowan} et~al.}{2008}]{McGowan2008}
{McGowan} K.~E.,  {Coe} M.~J.,  {Schurch} M.~P.~E.,  {Corbet} R.~H.~D.,
  {Galache} J.~L.,   {Udalski} A.,  2008, \mn@doi [\mnras]
  {10.1111/j.1365-2966.2007.12762.x}, \href
  {https://ui.adsabs.harvard.edu/abs/2008MNRAS.384..821M} {384, 821}

\bibitem[\protect\citeauthoryear{{Negueruela}, {Reig}, {Coe}  \&
  {Fabregat}}{{Negueruela} et~al.}{1998}]{Negueruela1998}
{Negueruela} I.,  {Reig} P.,  {Coe} M.~J.,   {Fabregat} J.,  1998, A\&A, \href
  {http://adsabs.harvard.edu/abs/1998A%26A...336..251N} {336, 251}

\bibitem[\protect\citeauthoryear{{Nixon} \& {King}}{{Nixon} \&
  {King}}{2012}]{NK2012}
{Nixon} C.~J.,  {King} A.~R.,  2012, \mn@doi [MNRAS]
  {10.1111/j.1365-2966.2011.20377.x}, \href
  {http://adsabs.harvard.edu/abs/2012MNRAS.421.1201N} {421, 1201}

\bibitem[\protect\citeauthoryear{{Nixon} \& {King}}{{Nixon} \&
  {King}}{2016}]{Nixon2016}
{Nixon} C.,  {King} A.,  2016, in {Haardt} F.,  {Gorini} V.,  {Moschella} U.,
  {Treves} A.,   {Colpi} M.,  eds, , Vol.~905, Lecture Notes in Physics, Berlin
  Springer Verlag.
p.~45, \mn@doi{10.1007/978-3-319-19416-5\_2}

\bibitem[\protect\citeauthoryear{{Nixon} \& {Pringle}}{{Nixon} \&
  {Pringle}}{2020}]{Nixon2020}
{Nixon} C.~J.,  {Pringle} J.~E.,  2020, \mn@doi [\apjl]
  {10.3847/2041-8213/abd17e}, \href
  {https://ui.adsabs.harvard.edu/abs/2020ApJ...905L..29N} {905, L29}

\bibitem[\protect\citeauthoryear{{Nixon} \& {Pringle}}{{Nixon} \&
  {Pringle}}{2021}]{Nixon2021}
{Nixon} C.~J.,  {Pringle} J.~E.,  2021, \mn@doi [\na]
  {10.1016/j.newast.2020.101493}, \href
  {https://ui.adsabs.harvard.edu/abs/2021NewA...8501493N} {85, 101493}

\bibitem[\protect\citeauthoryear{{Nixon}, {King}  \& {Price}}{{Nixon}
  et~al.}{2013}]{Nixonetal2013}
{Nixon} C.,  {King} A.,   {Price} D.,  2013, \mn@doi [MNRAS]
  {10.1093/mnras/stt1136}, \href
  {http://adsabs.harvard.edu/abs/2013MNRAS.434.1946N} {434, 1946}

\bibitem[\protect\citeauthoryear{{Ogilvie}}{{Ogilvie}}{1999}]{Ogilvie1999}
{Ogilvie} G.~I.,  1999, \mn@doi [MNRAS] {10.1046/j.1365-8711.1999.02340.x},
  \href {http://adsabs.harvard.edu/abs/1999MNRAS.304..557O} {304, 557}

\bibitem[\protect\citeauthoryear{{Ogilvie}}{{Ogilvie}}{2001}]{Ogilvie2001}
{Ogilvie} G.~I.,  2001, \mn@doi [MNRAS] {10.1046/j.1365-8711.2001.04416.x},
  \href {http://adsabs.harvard.edu/abs/2001MNRAS.325..231O} {325, 231}

\bibitem[\protect\citeauthoryear{{Ogilvie} \& {Dubus}}{{Ogilvie} \&
  {Dubus}}{2001}]{Ogilvie2001b}
{Ogilvie} G.~I.,  {Dubus} G.,  2001, \mn@doi [\mnras]
  {10.1046/j.1365-8711.2001.04011.x}, \href
  {https://ui.adsabs.harvard.edu/abs/2001MNRAS.320..485O} {320, 485}

\bibitem[\protect\citeauthoryear{{Okazaki} \& {Negueruela}}{{Okazaki} \&
  {Negueruela}}{2001}]{Okazaki2001}
{Okazaki} A.~T.,  {Negueruela} I.,  2001, \mn@doi [A\&A]
  {10.1051/0004-6361:20011083}, \href
  {http://adsabs.harvard.edu/abs/2001A%26A...377..161O} {377, 161}

\bibitem[\protect\citeauthoryear{{Okazaki}, {Bate}, {Ogilvie}  \&
  {Pringle}}{{Okazaki} et~al.}{2002}]{Okazaki2002}
{Okazaki} A.~T.,  {Bate} M.~R.,  {Ogilvie} G.~I.,   {Pringle} J.~E.,  2002,
  \mn@doi [MNRAS] {10.1046/j.1365-8711.2002.05960.x}, \href
  {http://adsabs.harvard.edu/abs/2002MNRAS.337..967O} {337, 967}

\bibitem[\protect\citeauthoryear{{Papaloizou} \& {Lin}}{{Papaloizou} \&
  {Lin}}{1995}]{PL1995}
{Papaloizou} J.~C.~B.,  {Lin} D.~N.~C.,  1995, \mn@doi [ApJ] {10.1086/175127},
  \href {http://adsabs.harvard.edu/abs/1995ApJ...438..841P} {438, 841}

\bibitem[\protect\citeauthoryear{{Papaloizou} \& {Pringle}}{{Papaloizou} \&
  {Pringle}}{1983}]{Papaloizou1983}
{Papaloizou} J.~C.~B.,  {Pringle} J.~E.,  1983, \mn@doi [\mnras]
  {10.1093/mnras/202.4.1181}, \href
  {https://ui.adsabs.harvard.edu/abs/1983MNRAS.202.1181P} {202, 1181}

\bibitem[\protect\citeauthoryear{{Papaloizou} \& {Terquem}}{{Papaloizou} \&
  {Terquem}}{1995}]{Papaloizou1995}
{Papaloizou} J. C.~B.,  {Terquem} C.,  1995, \mn@doi [\mnras]
  {10.1093/mnras/274.4.987}, \href
  {https://ui.adsabs.harvard.edu/abs/1995MNRAS.274..987P} {274, 987}

\bibitem[\protect\citeauthoryear{{Porter}}{{Porter}}{1996}]{Porter1996}
{Porter} J.~M.,  1996, MNRAS, \href
  {http://adsabs.harvard.edu/abs/1996MNRAS.280L..31P} {280, L31}

\bibitem[\protect\citeauthoryear{{Porter} \& {Rivinius}}{{Porter} \&
  {Rivinius}}{2003}]{Porter2003}
{Porter} J.~M.,  {Rivinius} T.,  2003, \mn@doi [\pasp] {10.1086/378307}, \href
  {http://adsabs.harvard.edu/abs/2003PASP..115.1153P} {115, 1153}

\bibitem[\protect\citeauthoryear{{Price}}{{Price}}{2007}]{Price2007}
{Price} D.~J.,  2007, \mn@doi [Pasa] {10.1071/AS07022}, \href
  {http://adsabs.harvard.edu/abs/2007PASA...24..159P} {24, 159}

\bibitem[\protect\citeauthoryear{{Price}}{{Price}}{2012}]{Price2012a}
{Price} D.~J.,  2012, \mn@doi [Journal of Computational Physics]
  {10.1016/j.jcp.2010.12.011}, \href
  {http://adsabs.harvard.edu/abs/2012JCoPh.231..759P} {231, 759}

\bibitem[\protect\citeauthoryear{{Price} \& {Federrath}}{{Price} \&
  {Federrath}}{2010}]{Price2010}
{Price} D.~J.,  {Federrath} C.,  2010, \mn@doi [\mnras]
  {10.1111/j.1365-2966.2010.16810.x}, \href
  {https://ui.adsabs.harvard.edu/abs/2010MNRAS.406.1659P} {406, 1659}

\bibitem[\protect\citeauthoryear{{Price} et~al.,}{{Price}
  et~al.}{2018}]{Price2018}
{Price} D.~J.,  et~al., 2018, \mn@doi [\pasa] {10.1017/pasa.2018.25}, \href
  {http://adsabs.harvard.edu/abs/2018PASA...35...31P} {35, e031}

\bibitem[\protect\citeauthoryear{{Pringle}}{{Pringle}}{1991}]{Pringle1991}
{Pringle} J.~E.,  1991, MNRAS, \href
  {http://adsabs.harvard.edu/abs/1991MNRAS.248..754P} {248, 754}

\bibitem[\protect\citeauthoryear{{Rajoelimanana}, {Charles}  \&
  {Udalski}}{{Rajoelimanana} et~al.}{2011}]{Rajoelimanana2011}
{Rajoelimanana} A.~F.,  {Charles} P.~A.,   {Udalski} A.,  2011, \mn@doi
  [\mnras] {10.1111/j.1365-2966.2011.18243.x}, \href
  {https://ui.adsabs.harvard.edu/abs/2011MNRAS.413.1600R} {413, 1600}

\bibitem[\protect\citeauthoryear{{Rajoelimanana}, {Charles}, {Meintjes},
  {Townsend}, {Schurch}  \& {Udalski}}{{Rajoelimanana}
  et~al.}{2017}]{Rajoelimanana2017}
{Rajoelimanana} A.~F.,  {Charles} P.~A.,  {Meintjes} P.~J.,  {Townsend} L.~J.,
  {Schurch} M.~P.~E.,   {Udalski} A.,  2017, \mn@doi [\mnras]
  {10.1093/mnras/stw2534}, \href
  {https://ui.adsabs.harvard.edu/abs/2017MNRAS.464.4133R} {464, 4133}

\bibitem[\protect\citeauthoryear{{Reig}, {Fabregat}  \& {Coe}}{{Reig}
  et~al.}{1997}]{Reig1997}
{Reig} P.,  {Fabregat} J.,   {Coe} M.~J.,  1997, \aap, \href
  {https://ui.adsabs.harvard.edu/abs/1997A&A...322..193R} {322, 193}

\bibitem[\protect\citeauthoryear{{R{\'{\i}}mulo} et~al.,}{{R{\'{\i}}mulo}
  et~al.}{2018}]{Rimulo2018}
{R{\'{\i}}mulo} L.~R.,  et~al., 2018, \mn@doi [\mnras] {10.1093/mnras/sty431},
  \href {http://adsabs.harvard.edu/abs/2018MNRAS.476.3555R} {476, 3555}

\bibitem[\protect\citeauthoryear{{Rivinius}, {Carciofi}  \&
  {Martayan}}{{Rivinius} et~al.}{2013}]{Rivinius2013}
{Rivinius} T.,  {Carciofi} A.~C.,   {Martayan} C.,  2013, \mn@doi [\aapr]
  {10.1007/s00159-013-0069-0}, \href
  {http://adsabs.harvard.edu/abs/2013A%26ARv..21...69R} {21, 69}

\bibitem[\protect\citeauthoryear{{Salvesen} \& {Pokawanvit}}{{Salvesen} \&
  {Pokawanvit}}{2020}]{Salvesen2020}
{Salvesen} G.,  {Pokawanvit} S.,  2020, \mn@doi [\mnras]
  {10.1093/mnras/staa1094}, \href
  {https://ui.adsabs.harvard.edu/abs/2020MNRAS.495.2179S} {495, 2179}

\bibitem[\protect\citeauthoryear{{Shakura} \& {Sunyaev}}{{Shakura} \&
  {Sunyaev}}{1973}]{SS1973}
{Shakura} N.~I.,  {Sunyaev} R.~A.,  1973, A\&A, \href
  {http://adsabs.harvard.edu/abs/1973A%26A....24..337S} {24, 337}

\bibitem[\protect\citeauthoryear{{Skinner}}{{Skinner}}{1980}]{Skinner1980}
{Skinner} G.~K.,  1980, \mn@doi [\nat] {10.1038/288141a0}, \href
  {https://ui.adsabs.harvard.edu/abs/1980Natur.288..141S} {288, 141}

\bibitem[\protect\citeauthoryear{{Skinner}}{{Skinner}}{1981}]{Skinner1981}
{Skinner} G.~K.,  1981, \mn@doi [\ssr] {10.1007/BF01246061}, \href
  {https://ui.adsabs.harvard.edu/abs/1981SSRv...30..441S} {30, 441}

\bibitem[\protect\citeauthoryear{{Slettebak}}{{Slettebak}}{1982}]{Slettebak1982}
{Slettebak} A.,  1982, \mn@doi [ApJs] {10.1086/190820}, \href
  {http://adsabs.harvard.edu/abs/1982ApJS...50...55S} {50, 55}

\bibitem[\protect\citeauthoryear{{Smallwood}, {Martin}  \& {Lubow}}{{Smallwood}
  et~al.}{2023}]{Smallwood2023}
{Smallwood} J.~L.,  {Martin} R.~G.,   {Lubow} S.~H.,  2023, \mn@doi [\mnras]
  {10.1093/mnras/stad338}, \href
  {https://ui.adsabs.harvard.edu/abs/2023MNRAS.520.2952S} {520, 2952}

\bibitem[\protect\citeauthoryear{{Suffak}, {Jones}  \& {Carciofi}}{{Suffak}
  et~al.}{2022}]{Suffak2022}
{Suffak} M.,  {Jones} C.~E.,   {Carciofi} A.~C.,  2022, \mn@doi [\mnras]
  {10.1093/mnras/stab3024}, \href
  {https://ui.adsabs.harvard.edu/abs/2022MNRAS.509..931S} {509, 931}

\bibitem[\protect\citeauthoryear{{Townsend} \& {Charles}}{{Townsend} \&
  {Charles}}{2020}]{Townsend2020}
{Townsend} L.~J.,  {Charles} P.~A.,  2020, \mn@doi [\mnras]
  {10.1093/mnrasl/slaa078}, \href
  {https://ui.adsabs.harvard.edu/abs/2020MNRAS.495L.139T} {495, 139}

\bibitem[\protect\citeauthoryear{{White} \& {Carpenter}}{{White} \&
  {Carpenter}}{1978}]{White1978}
{White} N.~E.,  {Carpenter} G.~F.,  1978, \mn@doi [\mnras]
  {10.1093/mnras/183.1.11P}, \href
  {https://ui.adsabs.harvard.edu/abs/1978MNRAS.183P..11W} {183, 11P}

\bibitem[\protect\citeauthoryear{{Zanazzi} \& {Lai}}{{Zanazzi} \&
  {Lai}}{2017}]{Zanazzi2017}
{Zanazzi} J.~J.,  {Lai} D.,  2017, \mn@doi [\mnras] {10.1093/mnras/stx208},
  \href {http://adsabs.harvard.edu/abs/2017MNRAS.467.1957Z} {467, 1957}

\bibitem[\protect\citeauthoryear{{Zanazzi} \& {Lai}}{{Zanazzi} \&
  {Lai}}{2018}]{Zanazzi2018}
{Zanazzi} J.~J.,  {Lai} D.,  2018, \mn@doi [\mnras] {10.1093/mnras/stx2375},
  \href {http://adsabs.harvard.edu/abs/2018MNRAS.473..603Z} {473, 603}

\bibitem[\protect\citeauthoryear{{von Zeipel}}{{von
  Zeipel}}{1910}]{vonZeipel1910}
{von Zeipel} H.,  1910, \mn@doi [Astronomische Nachrichten]
  {10.1002/asna.19091832202}, \href
  {https://ui.adsabs.harvard.edu/abs/1910AN....183..345V} {183, 345}

\makeatother
\end{thebibliography}








\bsp	
\label{lastpage}
\end{document}